\documentclass[11pt]{article}
\usepackage[usenames,dvipsnames]{color}
\usepackage{amsfonts}
\usepackage{amsmath}
\usepackage{amssymb}
\usepackage{graphicx}

\numberwithin{equation}{section}

\def\be{\begin{equation}}
\def\ee{\end{equation}}
\def\bea{\begin{eqnarray}}
\def\eea{\end{eqnarray}}

\newcommand{\eq}[1]{(\ref{#1})}

\def \ch{{\cal H}} 
\def \dc{\delta_c}
\def \dx{\delta_x}
\def \cshat{\hat{c_s}^2}
\def \cad{{c_a}^2}
\def \cs{{c_s}^2}
\def \fac{{\cal F}}
\def \dac{{\cal D}}

\begin{document}

\setcounter{footnote}{0}
\setcounter{figure}{0}
\setcounter{table}{0}

\vspace{-0.5cm}
\thispagestyle{empty}
\begin{flushright}
{\small CERN-PH-TH/2010-092\\ 
LAPTH-017-10}
 \end{flushright}
 
\vspace{1.0cm}
 
\begin{center}

{\Large {\bf Dark energy with non-adiabatic sound speed:\\ initial conditions and detectability}}
\vspace{0.7cm}
\end{center}
\vspace{0.23cm}
\begin{center}
\textbf{Guillermo Ballesteros\footnote{ballesteros@pd.infn.it}}\\ 
%{\small \href{mailto:ballesteros@pd.infn.it}{ballesteros@pd.infn.it}}  \\ 
\vspace{0.25cm}
 {\it \footnotesize  Museo Storico della Fisica e Centro Studi e Ricerche ``Enrico Fermi''.
Piazza del Viminale 1, I-00184, Rome.}\\ {\it \footnotesize  Dipartimento di Fisica ``G. Galilei'', Universit\`a degli Studi di Padova, via Marzolo
8, I-35131 Padova, Italy.}\\ {\it \small  INFN, Sezione di Padova, via Marzolo 8, I-35131 Padua, Italy.} \\
\vspace{0.5cm}
\textbf{Julien Lesgourgues\footnote{julien.lesgourgues@cern.ch}}\\ 
%{\small \href{julien.lesgourgues@cern.ch}{julien.lesgourgues@cern.ch}} \\ 
\vspace{0.25cm}
\textit{\footnotesize CERN, Theory Division, CH-1211 Geneva 23, Switzerland.}\\ 
\textit{\footnotesize Institut de Th\'eorie des Ph\'enom\`enes Physiques, EPFL, CH-1015 Lausanne,
Switzerland.}\\ 
\textit{\footnotesize LAPTh (CNRS - Universit\'e de Savoie), BP 110, F-74941 Annecy-le-Vieux
Cedex, France.}
\end{center}

\vspace{0.5cm}

\begin{abstract}
  Assuming that the universe contains a dark energy fluid with a
  constant linear equation of state and a constant sound speed, we
  study the prospects of detecting dark energy perturbations using CMB
  data from Planck, cross-correlated with galaxy distribution maps
  from a survey like LSST.  We update previous estimates by carrying a full exploration of the mock data likelihood for key fiducial models. We find that it will only
    be possible to exclude values of the sound speed very close to
    zero, while Planck data alone is not powerful enough for achieving
    any detection, even with lensing extraction.  We also discuss the
    issue of initial conditions for dark energy perturbations in the
    radiation and matter epochs, generalizing the usual adiabatic
    conditions to include the sound speed effect. However, for most
    purposes, the existence of attractor solutions renders the
    perturbation evolution nearly independent of these initial
    conditions.
\end{abstract}
\vspace{-0.2cm}

\newpage
\setcounter{page}{1}

\section{Introduction}

The biggest problem in cosmology today is the understanding of the
accelerated expansion of the universe. Although one could try to
attack this question leaving aside the cosmological principle or
modifying Einstein's gravity, the most classical approach consists of
assuming a perturbed FLRW universe with a negative pressure
component. The minimal model (in terms of number of free parameters)
compatible with the current data is a cosmological constant,
which should be perfectly homogeneous by definition.  Other candidates
(which may or may not alleviate the fine--tuning and coincidence
problems of the cosmological constant) include, for instance, scalar
field models, or effective descriptions in terms of a
fluid with free parameters yet to be measured. A canonically kinetic normalized scalar field would
fluctuate, but since in that case the sound speed $c_s^2 \equiv \delta
p / \delta \rho$ (computed in the rest frame of the scalar field) is
equal to one, local pressure would prevent density contrasts to grow
significantly. In an effective fluid description, the sound speed is a
free parameter, and dark energy clustering can be more efficient in
the limit in which overdensities are not balanced by local pressure
perturbations ($c_s \rightarrow 0$)\,.

Generally speaking, the study of small perturbations could be used as
a tool for discriminating between various models with a negative
pressure component (cosmological constant, dark energy fluid,
quintessence or k--essence fields, coupled dark energy, etc.) or
a modified theory of gravity. One of the major difficulties comes
from the fact that the expansion history predicted by a given
Lagrangian theory of gravity can be reproduced in General Relativity
by a dark fluid having an appropriate (time varying) equation of
state, or by a scalar field with an adequate kinetic term and
potential. Fortunately, a very precise measurement of clustering
properties in our universe could at least help to discard some models
in favor of others at the level of perturbations
\cite{Bertschinger:2008zb,Kunz:2006ca}. However, the spatial
fluctuations of typical dark energy models are very much suppressed
with respect to those of dark matter, and detecting their effect is a
real challenge.

The effects of quintessence perturbations (for which $c_s=1$) on the
CMB and LSS power spectra were discussed in \cite{DeDeo:2003te}. For
an (uncoupled) dark energy fluid, there have been several studies on
the possibility to measure $c_s\,$, but its value remains unconstrained
with present data (see e.g.
\cite{Bean:2003fb,Weller:2003hw,Hannestad:2005ak,Corasaniti:2005pq,Xia:2007km,Mota:2007sz,dePutter:2010vy,Li:2010ac}).
In a recent attempt, the authors
of~\cite{dePutter:2010vy} used present Cosmic Microwave Background (CMB), Large Scale Structure (LSS) and supernovae data
(including CMB$\times$LSS cross--correlation), and showed that it is
possible to see some weak preference for $c_s\neq 1\,$, but only for a
certain kind of early dark energy model in which the equation of
state is not constant.  We may still hope to discriminate between
different values of $c_s$ using combinations of future CMB data with
3--dimensional galaxy clustering data \cite{Takada:2006xs}, with
CMB$\times$LSS cross--correlation data \cite{Hu:2004yd}, or with
results from a large neutral hydrogen survey such as that of the SKA
project \cite{TorresRodriguez:2008et} (see also
\cite{TorresRodriguez:2007mk}).

In this work, we focus on an effective description which has already
been studied by several authors: namely, a dark energy fluid with a
linear equation of state $p=w\rho\,$, a constant equation of state
parameter $w$ close to $-1$ and a constant sound speed defined in the
range $0 \leq c_s^2 \leq1\,$.  In Section \ref{sspeed} we review the
concept of sound speed for a cosmological fluid. The
  purpose of Section \ref{initcond} is to clarify the non-trivial
  issue of initial conditions for dark energy perturbations in the
  radiation era, which are a priori non--adiabatic since $c_s^2 >
  w\,$. We present for the first time the initial conditions
  fulfilled by the dark energy fluid in the synchronous gauge (i.e.,
  in the gauge used by most Boltzmann codes), when all other fluids
  have adiabatic primordial perturbations. In Section \ref{attractors}
  we study analytically the evolution of these perturbations during
  the matter epoch. We derive approximations for the attractor
  solutions followed by dark energy perturbations (both in the
  Newtonian and synchronous gauges). These new results can be used in
  the future for analytical estimates of the impact of dark energy on
  structure formation. In Section \ref{detect}, in order to update the
  analysis of~\cite{Hu:2004yd}, we carry a full Monte-Carlo
  exploration of the likelihood of future mock CMB and LSS
  data, in order to infer the sensitivity of these data to the dark
energy sound speed, and to investigate possible parameter
degeneracies. Finally, in Section \ref{conclude}, we summarize our
findings.

\section{The sound speed} \label{sspeed}

At the level of inhomogeneities, the sound speed of a cosmological
fluid plays a similar role to that of the equation of state for the
background cosmology, and relates the pressure and density
perturbations as: 
\be \label{speed} \cs=\frac{\delta p}{\delta \rho}\,.
\ee 
Defined in this way, the sound speed is gauge dependent.
Indeed, the quantity that can be assumed to be a definite number
(depending on the microscopic properties of the fluid) is the
ratio $\delta p / \delta \rho$ evaluated in the fluid rest frame, often denoted
as $\cshat$. In another arbitrary frame, $\delta p / \delta \rho$ gets
corrections related to the velocity of the fluid in that frame, such that in Fourier space \cite{Bean:2003fb}
\be \label{restframe} \rho^{-1}\delta p = \cs \delta
= \cshat \delta + 3 \ch
\left(1+w\right)\left(\cshat-\cad\right)\frac{\theta}{k^2}\,, 
\ee
where $\delta=\delta\rho/\rho$ is the relative density perturbation,
$\theta$ the velocity divergence of the fluid, $k$ the comoving
wavenumber, $\ch$ the conformal Hubble parameter $d \ln a / d \tau$,
and $\cad$ the adiabatic sound speed of the fluid.  The latter is
defined as the (time dependent) proportionality coefficient between
the time variations of the background pressure and energy density of
the fluid, \be \dot p = \cad{\dot \rho}\,.  \ee For non--interacting
fluids one finds that \be \label{noni} \dot w =
3(1+w)\left(w-\cad\right)\ch\,, \ee which implies that $\cad=w$ if $w$
is constant and different from $-1$\,. In the rest of this work, we
will focus on the case in which $w$ can indeed be approximated as a
constant in time. Besides, we will restrict our analysis to
$0\leq\cshat\leq 1\,$. The lower bound prevents dark energy
fluctuations from growing exponentially, which can lead to unphysical
situations; and the upper one is imposed in order to avoid
superluminal propagation.  For a study on the implications of the sign
and value of $c_s^2$ for quintessence see \cite{Creminelli:2008wc}. For an expression relating the time evolution of $w$ and $c_s$ through the intrinsic entropy contribution to the pressure perturbation see \cite{Hu:1998kj} (and also \cite{Dent:2008ek}).

\section{Initial Conditions} \label{initcond}

In order to compute the CMB and LSS power spectra, one needs to evolve
cosmological perturbations starting from initial conditions deep
inside the radiation epoch and far outside the Hubble radius. Initial
conditions for photons, neutrinos, cold dark matter and baryons are
reviewed in \cite{Ma:1995ey} in the synchronous and Newtonian gauges
(see also \cite{Bucher:1999re}). Here, we want to extend
this set of relations to dark energy perturbations, especially in the
gauge used by most Boltzmann codes: namely, the synchronous
gauge. Surprinsingly, this issue has been overlooked in the
literature, without a clear justification\footnote{There were however various studies closely related to this
issue. For instance, in \cite{Dave:2002mn} various possible
initial conditions for the quintessence field, and their impact on the
CMB were discussed. In \cite{Doran:2003xq}, the initial conditions for a dark
energy fluid with quintessence-like perturbations were obtained in a gauge invariant
formalism. In \cite{Valiviita:2008iv}, the technique of
\cite{Doran:2003xq} was extended to interacting dark energy models. In all
these works, the dark energy sound speed $\hat{c}_s$ was kept fixed to
one.}. In practical terms, initial conditions for dark energy perturbations are essentially irrelevant for most purposes because of the existence of an attractor. We wish to clarify this issue and write down explicitly this attractor solution.

Adiabatic initial conditions on super--Hubble scale derive from
the generic assumption that for each component $i$, the density
$\rho_i(\tau,\vec{x})$ can be written as $\bar{\rho}_i(\tau + \delta
\tau(\vec{x}))\,$, where $\bar{\rho}_i(\tau)$ stands for the background density, and
$\delta \tau(\vec{x})$ is an initial time--shift function independent
of $i$. Similarly the pressure would read $\bar{p}_i(\tau + \delta
\tau(\vec{x}))$. Such conditions can be easily justified
in all models in which primordial perturbations are generated from a
single degree of freedom (like the inflaton), and/or in cases in which all components
have been in thermal equilibrium in the early universe with a common temperature and no
chemical potential. This form implies
\begin{eqnarray} \label{f1}
\rho_i &=& \bar{\rho}_i + \dot{\bar{\rho}}_i \, \delta \tau(\vec{x}),\\ \label{f2}
p_i &=& \bar{p}_i + \dot{\bar{p}}_i \, \delta \tau(\vec{x}),
\end{eqnarray}
and hence $\delta p_i / \delta \rho_i = \dot{\bar{p}}_i /
\dot{\bar{\rho}}_i\,$, i.e. the sound speed of each species must be
equal to its adiabatic sound speed. This generic assumption also
implies that the total ratio $[\sum_i \delta p_i] / [\sum_i \delta \rho_i]$ is independent
of the purely spatial coordinates $\vec{x}\,$. Finally, if all
the components do not interact, we conclude that 
\be
\frac{\delta_i(\tau,\vec{x})}{1+w_i} = - 3 \ch(\tau) \delta \tau(\vec{x})\,, \qquad \forall i\,.
\ee
The fact that for all species the ratios
  $\delta_i/(1+w_i)$ are equal to each other is a well-known property
  of adiabatic initial conditions. The meaning of such a relation is
  not so clear when one introduces a dark energy fluid, for which
  $\cshat > c_a^2$. Hence, in most frames, one has $c_s^2 \neq
  c_a^2\,$ and the fluid cannot obey
simultaneously (\ref{f1}) and (\ref{f2}). This
  raises the issue of defining sensible initial conditions for a
  dark energy fluid. However, during radiation and matter domination,
dark energy perturbations tend to fall inside the gravitational
potential wells created by the dominant component and not much concern
has been raised concerning their initial conditions. In other
words, there is an attractor solution for dark energy perturbations,
and their initial values are almost irrelevant in practice,
provided that for each Fourier mode the attractor is reached before
dark energy comes to dominate (i.e. provided that initial conditions
are imposed early enough, and that initial dark energy perturbations
are not set to dramatically large values). For this reason, in a
Boltzmann code like {\sc camb} \cite{Lewis:1999bs}, initial dark
energy perturbations are set by default to zero.

%In order to illustrate and clarify this issue, 
We will derive in the next subsections the attractor solution for a
dark fluid with constant $w$ and arbitrary $\hat{c}_s\,$, assuming
that other quantitites obey the usual adiabatic initial conditions,
and in two gauges: the Newtonian and the synchronous ones.
%. When
%  studying the evolution of cosmological perturbations, this solution
%  can be used to set initial conditions for the dark energy fluid, in
%  order to be sure that final results are independent of the time at
%  which initial conditions are defined.
%
%These initial conditions below cannot be called ``adiabatic'' because they do
%not correspond to $\delta p_i / \delta \rho_i = \dot{\bar{p}}_i /
%\dot{\bar{\rho}}_i$ for the dark energy fluid. However they stand for
%the correct solution for dark energy {\it when the other components obey
%adiabatic initial conditions}. Hence they could be called
%``generalized initial adiabatic conditions''. 
%
%Let us now derive this attractor solution for a dark energy
%fluid characterized by its equation of state $w$ and sound speed
%$\hat{c}_s\,$. 
For a detailed account of the construction of the two gauges
and the relations among them we refer the reader to
\cite{Ma:1995ey}\,. In what follows, we will denote quantities
corresponding to the conformal Newtonian gauge with a superscript. 
For instance, $\delta_x^{(c)}$ denotes the relative dark energy density in that gauge. No superscript will be used for quantities in the synchronous gauge. The transformation equations between the two gauges will be summarized below, later in this section.

\subsection{Synchronous gauge}

Early in the radiation era, the total energy density of the universe
can be approximated by the sum of photon and neutrino densities, with
a constant ratio
$R_\nu=\bar{\rho_\nu}/\left(\bar{\rho_\nu}+\bar{\rho_\gamma}\right)\,$.
In order to find the perturbation evolution on
  super-Hubble scales and for adiabatic initial conditions, one can
  combine the Einstein, photon and neutrino equations into a fourth
  order linear differential equation for the trace part $h\,$ of the
  metric perturbations in Fourier space \cite{Ma:1995ey}. The fastest
growing mode among the four possible solutions, $h\sim(k\tau)^2\,$,
corresponds to the growing adiabatic mode.

Since these conditions are established when the perturbations are
still in the super--Hubble regime, the product $(k\tau) \ll 1$ can be
used as an expansion parameter for the solutions of the dynamical
equations. At leading order
\bea 
\label{adiabinit} 
-\frac{1}{2} h =- \frac{1}{2} C(k\tau)^2 =
\dc = \delta_b =
\frac{3}{4}\delta_\nu =\frac{3}{4}\delta_\gamma\,, \eea 
where the subscripts refer to
cold dark matter, baryons, neutrinos and photons; and $C$ is a constant.
As usual, in order to fully fix the gauge, we impose not only
synchronous metric perturbations, but also that
dark matter particles have a vanishing velocity divergence $\theta_c\,$.
The continuity and Euler equations for the dark energy fluid read
\bea \label{laprimsim}
     \dot\dx&=&-(1+w)\left(\theta+\frac{\dot
       h}{2}\right)-3(\cshat-w)\ch\dx-9(1+w)(\cshat-\cad)\ch^2\frac{\theta_x}{k^2}\,,\\
\label{fortheta}
\dot\theta_x&=&-(1-3\cshat)\ch\theta_x+\frac{\cshat
  k^2}{1+w}\dx-k^2\sigma_x\,. \eea 
These equations are very general, since the only underlying assumption
is that the fluid is non--interacting, and allow for the presence of shear stress
$\sigma_x\,$, non--adiabatic sound speed, and a time varying $w\,$.
From now on we assume that the fluid is shear free and has a constant equation of state.
If the energy density of dark energy at early
times is negligible, the solution for
the metric perturbation $h\,$ will not change.
In order to find the attractor solution, we just need to replace $\dot{h}\,$ in \eq{laprimsim} according to
\eq{adiabinit}, and solve equations \eq{laprimsim}, \eq{fortheta}. As expected,
we find that the solution of the homogeneous equation becomes negligible with time, while
$\delta_x$ and $\theta_x$ are driven to
\bea \label{initx} \delta_x&=&-\frac{C}{2}
(1+w)\frac{4-3\cshat}{4-6w+3\cshat}(k\tau)^2\,,\\ \label{initxtheta}
\theta_x&=&-\frac{C}{2} \frac{\cshat}{4-6w+3\cshat} (k\tau)^3 k\,,
\eea 
at lowest order in $(k\tau)\,$.

This attractor solution does not look like usual adiabatic initial conditions
because, in general, $c_s^2$ is different from
$c_a^2$ in the synchronous gauge. Hence, $\delta p_x / \delta \rho_x$ cannot be equal to $\dot{\bar{p}}_x / \dot{\bar{\rho}}_x$.
However, this solution gives the correct behavior of
dark energy perturbations {\it when the other components obey
adiabatic initial conditions}, once the attractor has been reached. Therefore they could be called
``generalized initial adiabatic conditions''.
These conditions are valid not only for dark energy fluids
($w<-1/3$) but also for any other fluid with constant $w$ and
$\sigma=0$\,. For instance, one can easily check that the usual
adiabatic initial conditions for matter and radiation can be recovered
from (\ref{initx}, \ref{initxtheta}) by choosing $w = \cshat$.

In a Boltzmann code like {\sc camb} \cite{Lewis:1999bs},
the quantitites $\delta_x$ and $\theta_x$ are fixed to zero at initial
time for simplicity. For most practical applications, this arbitrary
choice does not introduce any mistake in the final results, since
$\delta_x$ and $\theta_x$ are quickly driven to the attractor solutions
of eqs.~(\ref{initx}, \ref{initxtheta}). We illustrate this in
Figure~\ref{fig_ini}, for a very large wavelength mode. We suggest however to
implement eqs.~(\ref{initx}, \ref{initxtheta}) directly into {\sc camb}'s
initial condition routine (as we did in Section~\ref{detect} of this
work), since this is completely straightforward, and since it offers a
guarantee that final results are independent of the early time at which
initial conditions are defined.

\subsection{Conformal Newtonian Gauge}

The equations that give the evolution of a generic fluid in the
conformal Newtonian gauge are (see also \cite{dePutter:2010vy}) \bea
\label{firstc}
\dot\delta_x^{(c)}&=&-(1+w)\left(\theta_x^{(c)}-3\dot\phi\right)-3\left(\cshat-w\right)\ch\dx^{(c)}-9(1+w)(\cshat-\cad)\ch^2\frac{\theta_x^{(c)}}{k^2}\,,\\
\label{secondc}
\dot\theta_x^{(c)}&=&-\left(1-3\cshat\right)\ch\theta_x^{(c)}+\frac{\cshat
k^2}{1+w} \delta_x^{(c)}-k^2\sigma_x^{(c)}+k^2\psi\,.  \eea The metric
perturbations \bea \phi&=&\eta -\alpha\ch\,,\\
\psi&=&\dot\alpha+\alpha\ch\,, \eea can be obtained from those of the
synchronous gauge using \be 2k^2\alpha=\dot h + 6 \dot \eta\,, \ee
where $\eta$ is the traceless part of the metric scalar perturbation
in the synchronous gauge in Fourier space. Using these last equations
one can immediately check that the product $\alpha\ch$ has zero time
derivative, so that $ \phi$ and $\psi$ are time independent at lowest
order in $(k\tau)\,$.  One can either solve directly (\ref{firstc},
\ref{secondc}), or use our results (\ref{initx}, \ref{initxtheta}) for
the behavior of $\delta_x$ and $\theta_x$ in the synchronous gauge,
and perform the gauge transformations \bea \label{t1}
\delta_x^{(c)}&=&-3(1+w)\alpha\ch+\dx\,,\\ \label{t2}
\theta_x^{(c)}&=&\alpha k^2+\theta_x\,,\\ \sigma_x^{(c)}&=&\sigma_x\,,
\eea that are valid for non--interacting fluids. The two methods give
\bea \delta_x^{(c)}&=&-\frac{3}{2}(1+w)\psi+\dx\,,\\
\theta_x^{(c)}&=&\frac{1}{2}\psi(k\tau)k+\theta_x\,, \eea where \be
\psi=\frac{20}{15+4R_\nu}C \ee and \be
\phi=(1+\frac{2}{5}R_\nu)\psi-\frac{5+4R_\nu}{6(15+4R_\nu)}C(k\tau)^2\,.
\ee Notice that the leading contributions to the velocity and density
perturbations in the conformal Newtonian gauge are independent of the
speed of sound and such that \bea \frac{\delta_x^{(c)}}{1+w} =
-\frac{3}{2} \psi = \delta_c^{(c)} = \delta_b^{(c)} =
\frac{3}{4}\delta_\nu^{(c)} =\frac{3}{4}\delta_\gamma^{(c)}\,. \eea
Therefore, the usual adiabatic conditions are recovered, and $\cshat$ enters
only in the next order corrections\footnote{ Note that the gauge transformation law (\ref{t1}) implies
that
\begin{equation}
\frac{\delta_x^{(c)}}{1+w} - \frac{\delta_j^{(c)}}{1+w_j} = 
\frac{\delta_x}{1+w} - \frac{\delta_j}{1+w_j},
\end{equation}
for any fluid $j$ with constant equation of state $w_j$. Since
the usual adiabatic conditions hold at leading order in the Newtonian
gauge, one may naively infer from the above equality that they hold
also in the synchronous gauge.  This is not correct since on the
left-hand side, the two terms are dominated by order zero terms in a
$(k \tau)$ expansion, while on the right-hand side the leading terms
are of order two. Assuming that the fluid $j$ has an adiabatic sound
speed (like cold dark matter or photons), a full order-two calculation
of all the terms leads to
\begin{equation}
\frac{\delta_x^{(c)}}{1+w} - \frac{\delta_j^{(c)}}{1+w_j} = 
\frac{\delta_x}{1+w} - \frac{\delta_j}{1+w_j}=
C \left(\frac{3 (\hat{c}_s^2-w)}{4 - 6 w + 3 \hat{c}_s^2}\right) (k \tau)^2.
\end{equation}
The righ-hand side does not vanish since $(\hat{c}_s^2-w)$ is by
assumption strictly positive. Being of order two, this difference
contributes to the solutions at leading order in the synchronous
gauge, but only at next-to-leading order in the Newtonian gauge.}.
Indeed, the Newtonian gauge is the one in which, beyond the Hubble
scale, $c_s^2$ is equal to $c_a^2$ at leading order, even when $\cshat
\neq c_a^2$. This can be checked by keeping the dominant terms in
(\ref{t1}, \ref{t2}), and replacing $\delta_x$ and $\theta_x$ by these
values in \eq{restframe}: one gets $c_s^2\delta_x^{(c)} =
c_a^2\delta_x^{(c)} + {\cal O}(k \tau)^2$. So, in the Newtonian gauge
and on super-horizon scale, $\delta \bar{p} / \delta \bar{\rho}$ is
equal to $p/\rho$ for any fluid with contant $w$, and the common
intuition according to which super-Hubble fluctuations behave in the
same way as background quantitites is recovered.

\section{Late time attractors} \label{attractors}

In the previous section, we found the attractor solutions for dark
energy perturbations during radiation domination. Here we will derive
similar solutions during matter domination. These results can be used to provide
initial conditions for dark energy perturbations in problems in which
following the behavior of cosmological perturbations during radiation
domination is not relevant.

If we assume that the energy density of photons and (massless)
neutrinos is negligible deep inside matter domination, the
perturbations can be studied using a two--fluid approximation. One of
the fluids is formed by baryons and cold dark matter (which cannot be
distinguished from each other) and the other one is dark energy. This
description in terms of two components can be accurately used to study
the growth of matter perturbations up to the present day.
Mathematically, the problem consists of a system of six independent
equations with six variables: the density and velocity perturbations
of the two fluids plus the scalar metric perturbations. We will now
find the relevant growing modes of the perturbations in the two gauges
and conclude this section with some remarks concerning the initial
conditions for dark energy.

\subsection{Synchronous gauge}

In the synchronous gauge, since we consider that cold dark matter and
dark energy are shear free, the Einstein equations imply that the two
metric degrees of freedom $\eta$ and $h$ are related to each
other. Eliminating $\eta$ in terms of $h$, and expressing $h$ in terms of
$\dc$ (the continuity equation for cold dark matter gives $\dot h =-2
\dot\dc$\,)\,, we obtain a system of two reasonably short second order
differential equations that describe the evolution of density
fluctuations 
\cite{Ballesteros:2008qk}\,:
\bea \label{scf}
\ddot\dc&+&\ch{\dot
\dc}-\frac{3}{2}\ch^2\Omega_c\dc=\frac{3}{2}\ch^2\Omega_x\left[\left(1+3\cshat\right)\dx+9\left(1+w\right)\ch\left(\cshat-w\right)\frac{\theta_x}{k^2}\right]\,,\\ \nonumber \nonumber \label{deper}
\ddot\dx&+&\left[3\left(\cshat-w\right)\ch-\fac(k,\ch)\right]\dot\dx\\
\nonumber
&-&\frac{3}{2}\left(\cshat-w\right)\left[\left(1+3w\Omega_x-6\cshat\right)\ch^2
  +2 \fac(k,\ch)\ch -\frac{2}{3}\frac{\cshat}{\cshat-w}k^2\right]
\dx\\&=& (1+w) \ddot\dc -(1+w) \fac(k,\ch) \dot \dc\,, \eea
where
\be
\label{diver}
(1+w)\frac{\theta_x}{k^2}=\frac{1}{\dac(k,\ch)}\left(-\dot\dx+(1+w)\dot\dc-3\left(\cshat-w\right)\ch\dx\right)\,,
\ee
and
\bea
\dac(k,\ch)&=&k^2+9(\cshat-w)\ch^2\,,\\ \label{fac}\fac(k,\ch)&=&
-9\left(1+3w\Omega_x\right)\frac{\cshat-w}{\dac(k,\ch)}\ch^3-(1-3\cshat)\ch\,.
\eea

In the purely matter dominated epoch ($\Omega_c=1$, $\Omega_x=0$) the
equation that describes the evolution of matter perturbations is the
classical growth formula: \be
\ddot\dc+\ch\dot\dc-\frac{3}{2}\ch^2\Omega_c\dc=0\,.  \ee Its general
solution is a linear combination of a growing mode $\left(\dc\sim
a\right)$ and a decaying one $\left(\dc \sim
a^{-3/2}\right)\,$. 
In order to find the relevant attractor solution
for dark energy perturbations, one should keep the first of these two solutions
\be
\dc\propto{(k\tau)^{2}}\,. \ee 
The behavior of cosmological perturbations depends on their wavelength
as compared to the characteristic scales of the problem. In our case,
and in terms of comoving scales, there are two relevant quantitites:
the comoving Hubble scale $\ch^{-1}\,$ (which gives the order of
magnitude of the causal horizon associated to any process starting
after inflation), and the comoving sound horizon
$\ch_s^{-1}=\hat{c_s}\ch^{-1}\,$. Analytical
approximations for the evolution of dark energy perturbations can be obtained
in the three regimes defined by these two scales.

For super--Hubble perturbations with $k\ll\ch\,$, one can approximate \eq{fac}
as $\fac(k,\ch) \simeq \left(3\cshat-2\right)\ch\,$, and
the homogeneous part of \eq{deper} gives two decaying modes for
$\dx\,$. The growing solution, sourced by the dark matter
perturbations, is \be \label{suphor}
\dx=(1+w)\frac{5-6\cshat}{5-15w+9\cshat}\dc\,, \quad \quad k\ll\ch\,.
\ee If instead $\ch_s\gg k\gg\ch\,$, the perturbations are below the
Hubble scale but above the sound horizon, and $\fac(k,\ch) \simeq
\left(3\cshat-1\right)\ch\,$. As in the previous case, the growing
mode for the density perturbations of dark energy comes from the one
of dark matter: \be \label{supsound}
\dx=(1+w)\frac{1-2\cshat}{1-3w+\cshat}\dc\,, \quad \quad \ch_s\gg
k\gg\ch\,.  \ee These formulas agree with the results of \cite{Sapone:2009mb} (when transformed into the synchronous gauge) if we take the sound speed to be zero. 
The equations \eq{suphor} and \eq{supsound} can be
used to define initial conditions for a dark energy fluid during
matter domination, provided that at the initial time all relevant wavelengths
are still larger than the sound horizon.  Below
the sound horizon, it becomes more difficult to obtain simple analytic
approximations.
%with
%equation of state $w$ and speed of sound $\cshat$ that generalize the
%standard adiabatic ones. Setting $\cshat=w$ we recover in both cases
%the usual formula for adiabatic initial conditions.  
When $\ch_s\ll k\,$, we can still approximate $\fac(k,\ch)$ as in the
previous case, but the term proportional to $\cshat k^2$ in \eq{deper}
becomes dominant. One gets: \be \ddot\dx+(1-3w)\ch\dot\dx+\cshat
k^2\dx=\frac{3}{2}(1+w)(1-2\cshat)\ch^2\dc\,, \quad \quad \ch_s\ll
k\,.  \label{insiders} \ee The general solution of the homogeneous
equation goes as $\tau^{3w-1/2}$ with a multiplicative factor that is
a linear combination of the Bessel functions $J_{3w-1/2}(c_s k \tau)$
and $Y_{3w-1/2}(c_s k \tau)\,$. The particular solution for the
complete equation including $\delta_c$ can also be written in terms of
non--elementary functions. Dark energy perturbations are anyway
suppressed with respect to dark matter ones in this regime, since
below the sound horizon (i.e below the Jeans length of the fluid) the
pressure perturbation can resist the gravitational infall. Indeed,
equation (\ref{insiders}) implies that in the limit $k \rightarrow
\infty\,$, the dark energy density contrast $\delta_x$ should vanish.

\subsection{Conformal Newtonian gauge}

In the conformal Newtonian gauge, the complete equations for the
evolution of density perturbations, equivalent to \eq{scf} and
\eq{deper}, become longer, because metric perturbations cannot be
trivially replaced in terms of $\delta_c\,$. Having obtained the
solutions in the synchronous gauge, it is simpler to apply \eq{t1}
rather than solving the conformal Newtonian equations directly. In the
limit of pure matter domination, i.e. $\Omega_x=0\,$, one can easily
prove that \be \dot h = 2 k^2 \alpha\,, \ee and therefore \bea \label{conf1}
\delta_c^{(c)}&=&\left(1+3\frac{\ch^2}{k^2}\right)\dc\,,\\ \label{conf2} \delta_x^{(c)}&=&\dx+3(1+w)\frac{\ch^2}{k^2}\dc\,.
\eea These equations show that outside the Hubble radius, the solution
is driven as usual to \be \delta_x^{(c)} =(1+w) \delta_c^{(c)}\,.  \label{newt_md_super}\ee
The same equations also imply that in the regime $\ch_s\gg k\gg\ch\,$,
\eq{supsound} remains valid in the conformal Newtonian gauge
(relations valid inside the Hubble radius are expected to be gauge
independent). 

\subsection{Attractor solutions}

An important feature in the evolution of dark energy perturbations in the two gauges,
which is common to the three regions we have studied, is that initial conditions for the dark energy perturbations
are almost irrelevant for the evolution in the purely matter dominated
epoch.  During this period, dark energy fluctuations track matter
inhomogeneities.  Above the sound horizon, as soon as the attractor
solution of \eq{suphor} or \eq{supsound} is reached, the ratio
$\delta_x/\delta_c$ only depends on the sound speed and
equation of state of the dark energy fluid. Notice that in the newtonian gauge
the sound speed dependence actually disappears outside the Hubble scale.

\begin{figure}[t]
\begin{center}
\includegraphics[width=8cm,angle=-90]{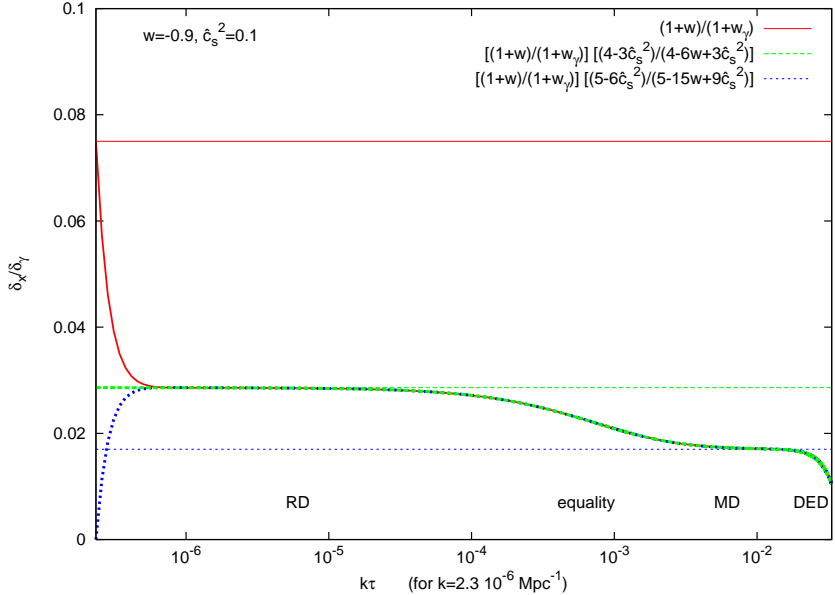}
\caption{ The three thick lines show the evolution of the ratio
  $\delta_x/\delta_\gamma$ in the synchronous gauge, obtained
  numerically with {\sc camb}, in a model with: $w=-0.9$, $\hat{c}_s^2
  = 0.1$, standard values of the other cosmological parameters, and
  adiabatic initial conditions for photons, neutrinos, cdm,
  baryons. We choose the case of a very long wavelength mode
  ($k=2.3\times 10^{-6}$Mpc$^{-1}$) which remains outside the Hubble
  radius during all relevant stages: radiation domination (RD), matter
  domination (MD) and dark energy domination (DED).  We integrated
  this mode starting either: {\it (i)} from the ``usual initial
  condition'' $\delta_x/\delta_\gamma=(1+w)/(1+w_\gamma)$ with
  $w_\gamma=1/3$ {\it (upper thin horizonal line)}, which has no physical
  justification in the synchronous gauge in this context; {\it (ii)}
  from eq.~(\ref{initx}) {\it (middle thin horizontal line)} and
  (\ref{initxtheta}); {\it (iii)} from $\delta_x=0$, like in the
  public version of {\sc camb}.  In each case, the solution quickly
  evolves in such way to fulfill eq.~(\ref{initx}) {\it (middle
    thin horizontal line)} during radiation domination, and then
  eq.~(\ref{suphor}) {\it (lower thin horizontal line)} during matter
  domination.
\label{fig_ini}}
\end{center}
\end{figure}

In the dark energy dominated period ($\Omega_x \rightarrow 1\,$), the
roles of the fluctuations are exchanged and matter perturbations are
sourced by dark energy ones. However, since we have seen that it is
only the initial value of $\delta_c$ that determines $\delta_x$ in the
matter epoch, the initial conditions for the evolution in the dark
energy period are in reality given only by $\delta_c\,$.  In
accordance with the results of section \ref{initcond}\,, this argument
can be extended back into the radiation era. Besides, from the
previous reasonings, it is clear that the velocity perturbations are
also unimportant. In conclusion, the amount of dark energy
perturbations today can be well estimated just by knowing the dark
matter perturbations at some initial time in the radiation or the
matter epoch. We illustrate this behavior in
  Figure~\ref{fig_ini} for a very large wavelength mode.

\section{Detectability} \label{detect} 

The detectability of the sound speed of dark energy has already been
studied by various authors, under different assumptions and for
various datasets. For example, for a model with constant $w$ and $c_s$
(identical to the one we consider in this work), the authors of
\cite{dePutter:2010vy} showed that the combination of present
CMB, LSS and
supernovae data is not sensitive at all to the dark energy sound
speed. Since dark energy perturbations would change the growth rate of
matter inhomogeneities on intermediate scales (between the Hubble
radius and the dark energy sound horizon), they can affect the CMB
photon temperature through the Late Integrated Sachs-Wolfe (LISW)
effect. Small variations in the LISW effect are difficult to detect in
the CMB temperature spectrum, due to cosmic variance and to the fact
that we only see the sum of primary anisotropies and LISW corrections.
However, the LISW contribution can be separated from primary
anisotropies by computing the statistical correlation between CMB and
projected LSS maps of the sky. In \cite{dePutter:2010vy}, most
of the presently available cross--correlation data were included in the
analysis, but current statistical and systematic error bars are far
too large for probing sub--dominant dark energy clustering effects.

One could think of constraining the dark energy sound
speed in the future, either by measuring with better accuracy the auto--correlation
function of matter distribution tracers (galaxy surveys, cluster
surveys, lensing surveys, \ldots) or, again, by studying their
cross--correlation with CMB maps in order to extract the LISW
contribution. The first option was studied for instance in
\cite{Takada:2006xs}, and the second in \cite{Hu:2004yd}.
In the latter, the authors focus on the cross--correlation of a CMB
experiment similar to Planck with a survey comparable to the Large
Synoptic Survey Telescope (LSST) project. The author of
\cite{Takada:2006xs} considered the combination of Planck
with future galaxy redshift surveys, not including any
cross--correlation information, but using the full three--dimensional
power spectrum of galaxies instead of their angular power spectrum
sampled in a few redshift bins (the loss of information on the LISW
effect can then be compensated by more statistics on the
auto-correlation function). Both works reached the same qualitative
conclusion that the next generation of LSS surveys could discriminate
between quintessence--like models with $c_s=1$ and alternative models
with a sound speed $c_s\ll 1$, and estimated under which threshold in
$c_s$ the discrimination would be significant.

The works of \cite{Takada:2006xs} and \cite{Hu:2004yd} are based on
analytic estimates of the measurement error for each cosmological
parameter, using the Fisher matrix of each future data set. This is
the second--order Taylor expansion with respect to the cosmological
parameters of the data likelihood around its maximum, i.e. in the
vicinity of a putative fiducial model. This method should be taken
with a grain of salt whenever the dependence of the observable
spectrum on a given parameter cannot be approximated at the linear
level within the range over which this parameter gives good fits to
the data. This is precisely the case for the model at hand. The dark
energy sound speed will not be pinned down with great accuracy using
the Fisher matrix approach. Its variation within the error bar has a
non--trivial effect on the total matter power spectrum which, although
small, cannot be captured at the linear level, since it amplifies
perturbations over a range of scales depending on $c_s$ itself. This
means that Fisher matrix estimates for the error $\Delta c_s$ should
only be considered as a first--order approximation. This caveat was
already emphasized in Section IV of \cite{Hu:2004yd}. Hence, it is
worth checking the results of \cite{Hu:2004yd} with a full
exploration of the likelihood of mock Planck + LSST
data. We generated such data for some fiducial models including a
dark energy fluid, sampled the data likelihood with a Monte Carlo
Markhov Chain (MCMC) approach, and inferred the marginalized posterior
probability distribution of the dark energy sound speed. Unlike the
Fisher matrix approach, this method takes into account the precise
effect of each parameter throughout the parameter space of the
model. Hence, it can assess realistically whether the data can resolve
non--trivial parameter degeneracies. As in \cite{Takada:2006xs}, we
included the total neutrino mass in the list of parameters to measure,
in order to check whether some confusion between the effect of dark
energy clustering and massive neutrino free--streaming could arise
when the dark energy sound horizon is comparable to the neutrino
free--streaming scale. Our mock data sets were generated and fitted
with two modified versions of the public {\sc CosmoMC}
package~\cite{Lewis:2002ah} which were already developed and described
by the authors of \cite{Perotto:2006rj} and~\cite{Lesgourgues:2007ix}.

\subsection{Planck with lensing extraction}

In a first step, we focus on the potential of the Planck experiment
alone, assuming Blue Book sensitivities for the most relevant
frequency channels, and a sky coverage of 65\% (as summarized in
\cite{Lesgourgues:2005yv}). Given that small variations in the LISW
effect cannot be probed due to cosmic variance, the only hope to
detect dark energy clustering with Planck data alone is through
lensing extraction. A lensing estimator can in principle measure the
deflection of CMB anisotropies caused by gravitational lenses (e.g.
galaxy clusters). This techniques allows to reconstruct the power
spectrum of the gravitational potential projected along the line of
sight, and to disentangle a fraction of the total LISW
contribution to temperature anisotropies (by cross--correlating the
lensing map with the temperature map). Like in
\cite{Perotto:2006rj}, we ran {\sc CosmoMC} in order to fit some
mock Planck data, including reconstructed lensing data with a noise
level corresponding to the ``minimum variance quadratic estimator'' of
\cite{Okamoto:2003zw}. The mock data generator and all
modifications to the {\sc CosmoMC} package are publicly
available\footnote{\tt http://lesgourg.web.cern.ch/lesgourg/codes.html}. Our
results show that Planck alone is completely insensitive to the dark
energy sound speed for any constant and reasonable value of
$w$. Indeed, when assuming a flat prior on $\log_{10}[c_s]$ in the
range $[-3,0]$, we obtained a nearly constant posterior distribution
${\cal P}(\log_{10}[c_s])$. We checked this results with several
assumptions on the fiducial values of $w$ (bewteen $-0.8$ and 1), $c_s$
(between $10^{-2}$ and 1) and $\Sigma m_\nu$ (between 0.05~eV and
0.2~eV). This negative result can be attributed, on the one hand, to
the limited precision of lensing extraction with Planck, and on the
other hand, to the fact that the projected gravitational potential
probed by CMB lensing gives more weight to high redshifts (of the
order of $z \sim 3$) than to the redshifts at which dark energy comes
to dominate and eventually affects the growth of total matter
perturbations ($z < 1$ for models with a constant $w$).  In the rest
of this section, we will not include Planck lensing extraction
data anymore.

\subsection{Planck plus LSST galaxy data}

\begin{figure}[t]
\begin{center}
\includegraphics[width=8cm]{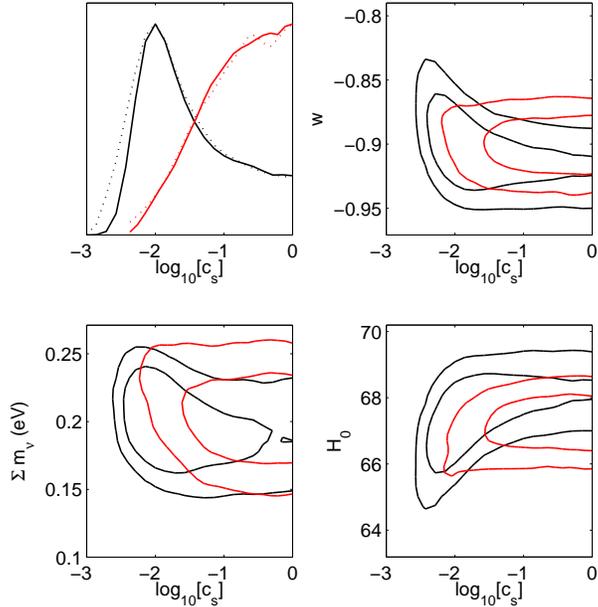}
\caption{{\it (Top left)} Marginalized posterior probability distribution of $\log_{10}[c_s]$
for Planck+LSST and the two fiducial models described in the text and in Table~\ref{table},
in which $c_s=1$ (red lines) or $c_s=10^{-2}$ (black lines). The dotted lines show the mean likelihood
for comparison. {\it (Other plots)} For the same models and mock data, joint 68\% and 95\% confidence
contours for $\log_{10}[c_s]$ and the three parameters most correlated to the dark energy sound speed.\label{figure}}\end{center}
\end{figure}

In a second step, we performed a combined analysis of mock Planck and
LSST data, including information on the angular auto--correlation
function of LSST galaxy maps (in 65\% of the sky and in 6 redshift
bins spanning the range from $z=0$ to $z \sim 5$), and on the
cross--correlation between these maps and CMB temperature maps.  Our
assumptions concerning LSST data binning, selection function, bias,
density and coverage are identical to those in
\cite{LoVerde:2006cj,Lesgourgues:2007ix}. The corresponding
expression for the likelihood of correlated CMB and galaxy data is
given in \cite{Lesgourgues:2007ix}.  We generated two mock data sets,
assuming two fiducial cosmological models with: standard values of the
six $\Lambda$CDM parameters, a total neutrino mass $\sum m_{\nu}
=0.2$~eV (equally split between the three species for simplicity),
a constant dark energy equation of state parameter $w=-0.9$, and a
dark energy sound speed equal to either $\hat{c_s}=1$ or $\hat{c_s}=10^{-2}$. For
each model we ran sixteen chains with flat priors on the usual {\sc
CosmoMC} parameter basis ($\omega_b$, $\omega_c$, $\theta$, $\tau$,
$\log[A_s]$, $n_s$, $f_{\nu}$, $w$) plus $\log_{10}[\hat{c_s}]$, imposing a
restriction $\log_{10}[c_s]\leq 0$.  After reaching a convergence
criterion $R-1 <0.01$, we obtained the marginalized posterior
probability of each of these parameters, and of the derived parameters
$\sum m_{\nu}$ (total neutrino mass) and $H_0$.  We show in
Table~\ref{table} the 68\% standard error for each of these. In the
first plot of Figure~\ref{figure}, we display the marginalized
distribution of $\log_{10}[\hat{c_s}]$ in the two models.  The correlation
between $\log_{10}[\hat{c_s}]$ and other parameters appears to be small,
except for $w$, $f_{\nu}$ (or $\sum m_{\nu}$), and $\theta$ (or
$H_0$). In Figure~\ref{figure}, we also show the two--dimensional 68\%
and 95\% confidence levels contours for these pairs of parameters.

\begin{table}
\begin{center}
\begin{tabular}{|l|cc|}
\hline
& \multicolumn{2}{c|}{Standard errors} \\
                      & $\hat{c_s}^{\rm fiducial}=1$ & $\hat{c_s}^{\rm fiducial}=10^{-2}$ \\
\hline
$\Omega_b h^2$        & 0.00011 & 0.00011 \\
$\Omega_c h^2$        & 0.00099 & 0.00095 \\
$\theta$              & 0.00024 & 0.00024 \\
$\tau$                & 0.0033 & 0.0028 \\
$\log_{10}[c_s]$      & $>$ {\it -1.1} & $>$ {\it -1.9} \\
$f_{\nu}$             & 0.0022 & 0.0021 \\
$w$                   & 0.017 & 0.021 \\
$n_s$                 & 0.0031 & 0.0030 \\
$\log[10^{10} A_s]$   & 0.0095 & 0.0090 \\
\hline
$\Sigma m_{\nu}$ (eV) & 0.024 & 0.023 \\
$H_0$ (km/s/Mpc)      & 0.64 & 0.87 \\
\hline
\end{tabular}
\end{center}
\caption{Standard deviation on each cosmological parameter for
  correlated CMB+galaxy data from Planck+LSST for two fiducial models
  with standard values of the six $\Lambda$CDM parameters, a total
  neutrino mass $\sum m_{\nu} =0.2$~eV (equally split between the
  three species), a constant dark energy equation of
  state $w=-0.9\,$ and a dark energy sound speed equal to
  either $c_s=1$ or $c_s=10^{-2}\,$.  The last two lines show derived
  parameters. For $\log_{10}[\hat{c_s}]$, instead of quoting the standard
  error, we provide the lower bound of the 68\% confidence interval (the upper
  bound being imposed by the prior, $\log_{10}[\hat{c_s}]\leq 0$). \label{table}}
\end{table}

We learn from these runs that correlated CMB+galaxy data from
Planck+LSST may discriminate between various sound speed
values with a standard deviation $\Delta \log_{10}[\hat{c_s}]$ of order 1\,. In other
words, such a data combination can give an indication on the order of
magnitude of $\hat{c_s}\,$. With a fiducial value $\hat{c_s}^{\rm fiducial}=1\,$, the
lower bound of the 68\% (resp. 95\%) confidence interval is
$\log_{10}[\hat{c_s}]>-1.1$ (resp. $-1.8$)\,. With a fiducial value $\hat{c_s}^{\rm
  fiducial}=10^{-2}$, these numbers become $\log_{10}[\hat{c_s}]>-1.9$
(resp. $-2.3$)\,. The 68\% confidence intervals always include
$\log_{10}[\hat{c_s}]=0$, although the case of $\hat{c_s}^{\rm fiducial}=10^{-2}$
appears to be at the threshold below which the data would impose a
negative 68\% upper bound on $\log_{10}[c_s]$.  These estimates are not as optimistic as those in \cite{Hu:2004yd}, most probably because
of our more accurate technique for error forecast (based on a
marginalization of the actual likelihood over the full parameter
space). They are also less dramatic than those of \cite{Takada:2006xs},
inferred from the Fisher matrix of a different dataset. Still, they
show in a robust way that with Planck+LSST one can exclude dark energy
models with maximum clustering (i.e. $\hat{c_s} \rightarrow 0$), which is
not the case with current data (and will not be the case even in one year from
now with Planck data). This limitation is due to the smallness of the
effect and not to parameter degeneracies, since $\log_{10}[\hat{c_s}]$ is
not particularly correlated with other parameters. Of course, the
higher is $w$, the longer will be the dark energy domination and the stronger
will be the effect of $\hat{c_s}$. This explains the correlation between
$\log_{10}[\hat{c_s}]=0$ and $w$ seen in Figure~\ref{figure}: if $w$ were
found to be closer to $-0.8$ than to $-1$, the measurement of $c_s$ would be
considerably easier. This is in agreement with the $c_s$ dependence of the bounds on $w$ found in \cite{Pietrobon:2006gh}. The joint ($\log_{10}[\hat{c_s}]$, $w$) likelihood
contours actually suggest that even with $w^{\rm fiducial}=-0.85\,$, a value like
$\hat{c_s}=10^{-2}$ could be pinned down with a fairly good accuracy. Like
in \cite{Takada:2006xs}, we find that the degeneracy between the
dark energy sound speed and the neutrino mass is not significant: this
can be understood from Figure 5. of \cite{Takada:2006xs}, which
shows that dark energy clustering and neutrino free--streaming can
produce a step in the matter power spectrum on similar scales, but
with different shape. The step induced by dark energy is
sharper because this effect occurs during a brief period of time
compared to non--relativistic neutrino free--streaming.

\section{Conclusions} \label{conclude}

In sections \ref{initcond} and \ref{attractors}, we have
  studied analytically the behavior of dark energy perturbations in
  the radiation and matter epochs in the synchronous and conformal
  Newtonian gauges. We have written down the formulas that generalize the usual
  adiabatic initial conditions for small inhomogeneities when a dark
  energy fluid with constant equation of state and speed of sound is
  considered. Our equations~(\ref{initx}, \ref{initxtheta}) can be
  readily implemented in the initial conditions of e.g. {\sc
    camb}, which applies to early radiation
  domination. Instead, equations~(\ref{suphor}, \ref{supsound}) -- or
  their counterpart (\ref{newt_md_super}, \ref{supsound}) in the
  Newtonian gauge -- provide correct initial conditions above the
  sound horizon for codes simulating only the growth of matter
  fluctuations during radiation and dark energy domination. The fact
  that there is an attractor behavior for the evolution of the
  perturbations guarantees that dark energy fluctuations are driven to
  solutions dictated by the dark matter ones. This implies that the
  final state of dark energy today is independent on its possible
  initial conditions, provided that the attractor solution is reached
  way before dark energy comes to dominate the background expansion;
  using the above set of initial conditions guarantees that such a
  condition is fullfilled.

After clarifying these issues, we have studied in section \ref{detect} the
prospects for detecting dark energy perturbations using CMB data from
the Planck satellite, cross-correlated with galaxy distribution maps
obtained with an LSST-like instrument.  We have chosen to focus on
models of dark energy consisting in a fluid with constant equation of
state $w$ and constant sound speed $\hat{c}_s$. We have performed a
full exploration of the likelihood of our mock data using a
  Monte Carlo approach, in order to cross-check previous Fisher matrix
  estimates in a more robust way. We find that the 
  confidence interval for $\hat{c}_s$ inferred from the data will
  potentially allow us to put a lower bound on $\hat{c}_s$, and to
  constrain at least its
  order of magnitude (although with a slightly
  smaller significance than in earlier analytic estimates).  Our
  results build upon previous works on the same topic
  \cite{Bean:2003fb,Hu:2004yd,Takada:2006xs,dePutter:2010vy} and
  complement other researches were different models have been studied
  (see for instance \cite{dePutter:2010vy} for early dark energy and
  \cite{Martinelli:2010rt} for the case in which there is a coupling
  to dark matter).

\section{Acknowledgments}
We thank Wessel Valkenburg for his large contribution to the numerical
module \cite{Lesgourgues:2007ix} that we have employed in our
analysis.  We wish to thank Juan Garcia-Bellido, Katrine Skovbo and
Toni Riotto for very useful discussions.  Numerical computations were
performed on the MUST cluster at LAPP (CNRS \& Universit\'e de
Savoie). GB thanks the CERN TH unit for
hospitality and support and INFN for support. JL acknowledges support from the EU 6th Framework Marie
Curie Research and Training network `UniverseNet'
(MRTN-CT-2006-035863).

%\bibliography{pertrefs}

\end{document}